\documentstyle[aps,epsf,prl%
,twocolumn%
]{revtex}
\begin{document}
\draft
\title{Controlling friction}
\author{Franz-Josef Elmer}
\address{Institut f\"ur Physik, Universit\"at
   Basel, CH-4056 Basel, Switzerland}
\date{November 1997}
\maketitle
\begin{abstract}
Two different controlling methods are proposed to stabilize 
unstable continuous-sliding states of a dry-friction oscillator.
Both methods are based on a delayed-feedback mechanism well-known
for stabilizing periodic orbits in deterministic chaos. The feedback
variable is the elastic deformation. The control parameter is either the
sliding velocity or the normal force. We calculate analytically
stability boundaries in the space of control parameter and delay
time. Furthermore, we show that our methods are able to turn
stick-slip motion into continuous sliding. Controlling friction helps
to get a better understanding
of friction by measuring, e.g., velocity-weakening friction forces.
\end{abstract}
\pacs{PACS numbers: 03.20.+i, 46.30.Pa, 07.05.Dz}

\narrowtext

If one tries to move two contacting solid bodies laterally, one often
observes stick-slip motion due to dry friction (i.e., solid-solid
friction with or without lubricants) \cite{bow.54}. This motion is
characterized by a (more or less) periodic switching between sticking
(relative sliding velocity is zero) and slipping (relative sliding
velocity is on average much larger than the applied velocity). This stick-slip
motion is responsible for the everyday experience of singing violins
and squeaking doors. In most technological cases one wants to avoid
stick-slip motion because it leads to vibrations and wear. The goal
is to bring the system into the continuous sliding state, where the
relative sliding velocity is constant and does not oscillate.

\begin{figure}
\epsfxsize=80mm\epsffile{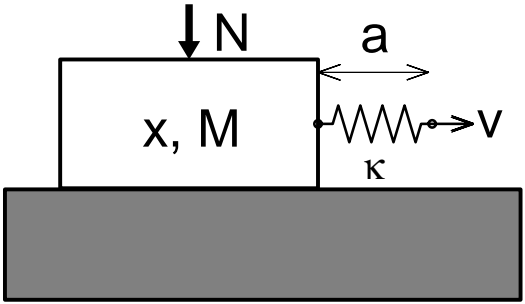}\vspace{3mm}
\caption[fsys]{\protect\label{f.sys}The stick-slip oscillator.
The feedback variable of the proposed controlling methods is the 
elastic deformation $a$. The control parameter is either the applied
velocity $v$ or the normal load $N$.
}
\end{figure}

The classical method of avoiding stick-slip motion is to use a
lubricant. But even at high normal load and small sliding velocities
stick-slip motion occurs \cite{bow.54}. Here we propose an active
way to avoid stick-slip motion. It is inspired by the methods used in 
controlling chaos \cite{ott.90,pyr.92}. The idea is not to change the 
physics at the friction interface but to stabilize unstable states. 

In our case the unstable state is the continuous-sliding state. It is
unstable for velocity-weakening friction laws where the friction
force decreases with increasing sliding velocity\cite{rem1}. We use 
the delayed-feedback method proposed
by Pyragas for stabilizing periodic orbits in a chaotic 
attractor\cite{pyr.92}. The feedback variable is the
elastic deformation (see below).  Our parameter of controlling is either
the sliding velocity or the normal load. 

In order to show that our proposed methods work, we have studied
analytically as well as numerically a simplified model for the
lateral motion of two solid bodies in contact (see Fig.~\ref{f.sys}).
We assume that one of the bodies is fixed whereas the other one (with
mass $M$) can slide. The elasticity of the sliding bodies (or the
whole machinery) is modeled by a spring with stiffness $\kappa$.
There is a normal load $N$ which presses the bodies against each
other and there is the applied velocity $v$ which is generally
different from the relative sliding velocity $\dot x$ of the bodies.
We assume that the lateral degree of freedom $x$ is the most
important one. We therefore neglect all other ones. The friction force
$F$ at the interface is proportional to $N$ in accordance with 
Amononton's law \cite{bow.54}. The equations of motion read
\begin{mathletters}\label{eqm}
\begin{equation}
  M\ddot x(t)=\kappa a(t)-\mu_K\bigl(\dot x(t)\bigr)N,\quad
   \mbox{if $\dot x(t)\neq 0$},
  \label{eqm.x1}
\end{equation}
or
\begin{equation}
  \dot x(t)=0,\quad\mbox{if $|\kappa a(t)|<\mu_SN$},
  \label{eqm.x2}
\end{equation}
and
\begin{equation}
  \dot a(t)=v-\dot x(t),
  \label{eqm.a}
\end{equation}
\end{mathletters}
where $\mu_S$ is static friction coefficient and $\mu_K(\dot x)$ is
the kinetic friction coefficient which in general depends on the
sliding velocity. The variable $a$ denotes the spring elongation
(i.e., the difference between the stage position and the block
position). In general it measures the elastic deformation of the
bodies due to the contact. Our feedback variable is $a$. The control
parameter is either the applied velocity $v$ or the normal load $N$.
Hence we replace either $v$ by
\begin{mathletters}\label{ctrl}
\begin{equation}
  v=v_0+\alpha_v[a(t)-a(t-\tau)]
  \label{ctrl.v}
\end{equation}
or $N$ by
\begin{equation}
  N=N_0+\alpha_N[a(t)-a(t-\tau)],
  \label{ctrl.N}
\end{equation}
\end{mathletters}
where $v_0$ and $N_0$ are the unperturbed applied velocity and normal
load, respectively, $\tau$ is the delay time, and $\alpha_v$ and
$\alpha_N$ are the amplitudes of control. Note, that load control
works only if $N$ is always positive. Otherwise it would lead to a
lift-off of the sliding bodies. We will first calculate analytically
where in the space of control amplitudes and delay time the
continuous-sliding state is stable. 

The continuous-sliding state is given by $\dot x=v_0$ and
$a=\mu_K(v_0)N_0/\kappa$. In order to test its stability we make the
ansatz $\dot x(t)=v_0+c_{\dot x}\exp(\lambda t)$ and
$a(t)=\mu_K(v_0)N_0/\kappa+c_a\exp(\lambda t)$ and linearize the 
equation of motion in $c_{\dot x}$ and $c_a$. Nontrivial solutions
are possible only if $\lambda$ fulfills the following characteristic
``polynomial'':
\begin{eqnarray}
\lefteqn{M\lambda^2+[\mu_K'(v_0)N_0-M\alpha_v(1-e^{-\lambda\tau})]
   \lambda+\kappa}\nonumber\\
  &&\hspace{10mm} -[\mu_K'(v_0)N_0\alpha_v+\mu_K(v_0)\alpha_N]
   (1-e^{-\lambda\tau})=0.
  \label{cp}
\end{eqnarray}

The delay term is responsible for $e^{-\lambda\tau}$ which turns the
polynomial into a transcendental equation for $\lambda$. Thus
(\ref{cp}) has more than two solutions. In fact there are infinitely
many. Solutions are either real or coming as conjugated
complex pairs. The continuous sliding state is stable if the real
parts of all solutions of (\ref{cp}) are negative. It can be shown
that the number of solutions with positive real part is finite
because the exponential term is bounded. In fact one can give upper
limits of the real part and the imaginary part of the solutions.
Actually we do not need to known the solutions of (\ref{cp}). We only
want to know the stability boundary in the space of the control
amplitudes $\alpha_v$, $\alpha_N$ and the delay time $\tau$. The
stability boundary is a two-dimensional manifold where a solution of
(\ref{cp}) crosses the imaginary axis.  Such manifolds can be
calculated analytically in parametric form.  They are solutions of
\begin{equation}
  2A_rB_r+2A_iB_i=B_r^2+B_i^2,
  \label{ac}
\end{equation}
\begin{mathletters}\label{tc}
\begin{equation}
  \cos\omega\tau=1-\frac{A_rB_r+A_iB_i}{A_r^2+A_i^2}
  \label{tc.c}
\end{equation}
and
\begin{equation}
  \sin\omega\tau=\frac{A_rB_i-A_iB_r}{A_r^2+A_i^2},
  \label{tc.s}
\end{equation}
\end{mathletters}
where
\begin{equation}
  A_r\equiv\mu_K(v_0)\alpha_N+\mu_K'(v_0)N_0\alpha_v,\quad
  A_i\equiv M\omega\alpha_v,
  \label{ar.ai}
\end{equation}
and
\begin{equation}
  B_r=\kappa-M\omega^2,\quad B_i=\mu_K'(v_0)N_0\omega.
  \label{br.bi}
\end{equation}
The parameter $\omega$ is the imaginary part of $\lambda$. 
Eq.~(\ref{ac}) is a linear inhomogeneous equation for $\alpha_v$ and
$\alpha_N$. By giving one $\alpha$ we can express the other one in
terms of $\omega$.  From (\ref{tc}) we get a countable set of
solutions for $\tau$. Near the manifold one can expand $\lambda$ into
a Taylor series in order to find out whether the real part of
$\lambda$ increases or decreases when the manifold is crossed.
Together with the facts that (\ref{cp}) has analytic solutions for
$\alpha_v=\alpha_N=0$ and the solutions of (\ref{cp}) are continuous
functions of the parameters we are able to find the stable
regions in the parameter space.

\begin{figure}
\epsfxsize=80mm\epsffile{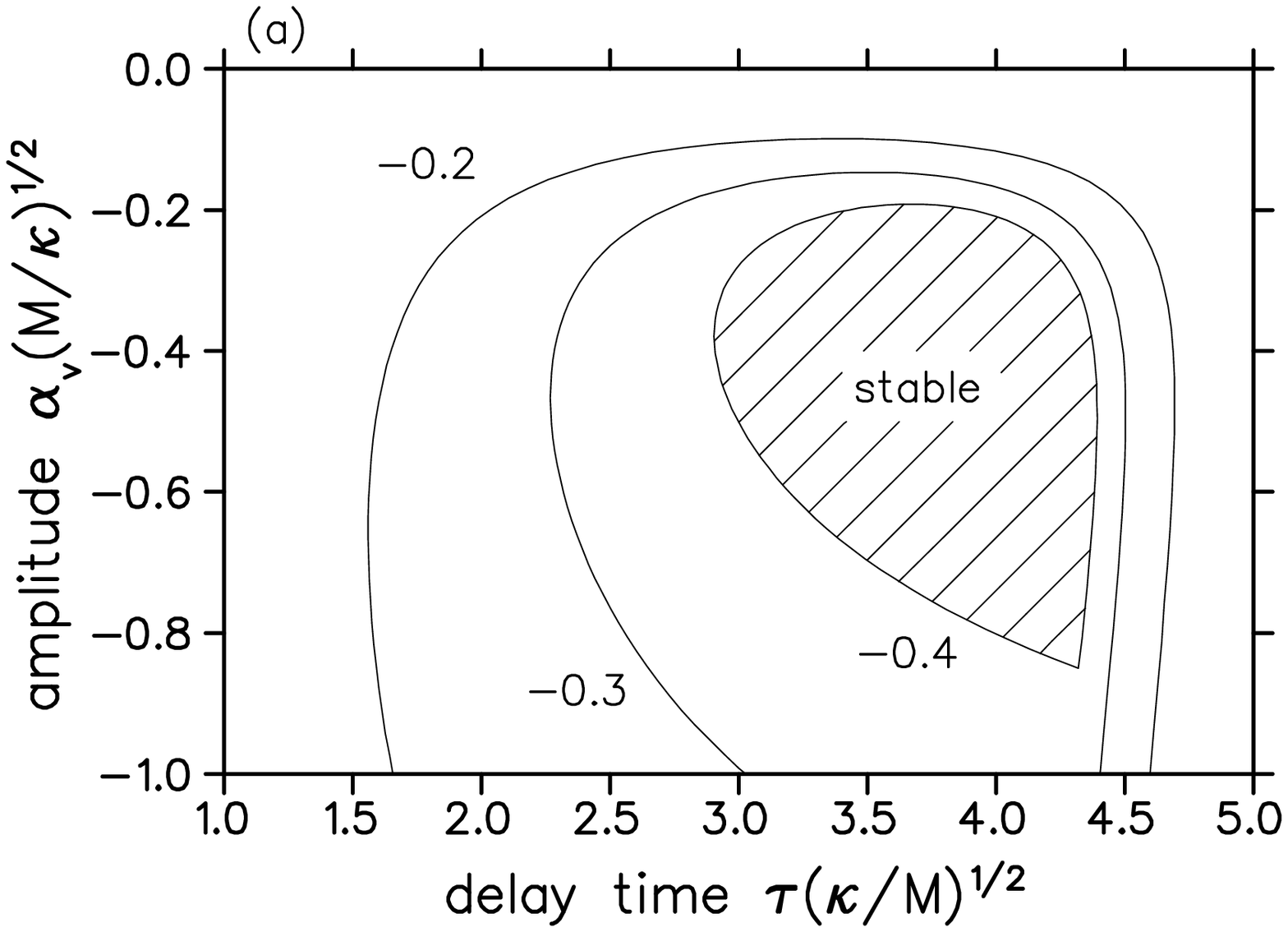}\vspace{3mm}
\epsfxsize=80mm\epsffile{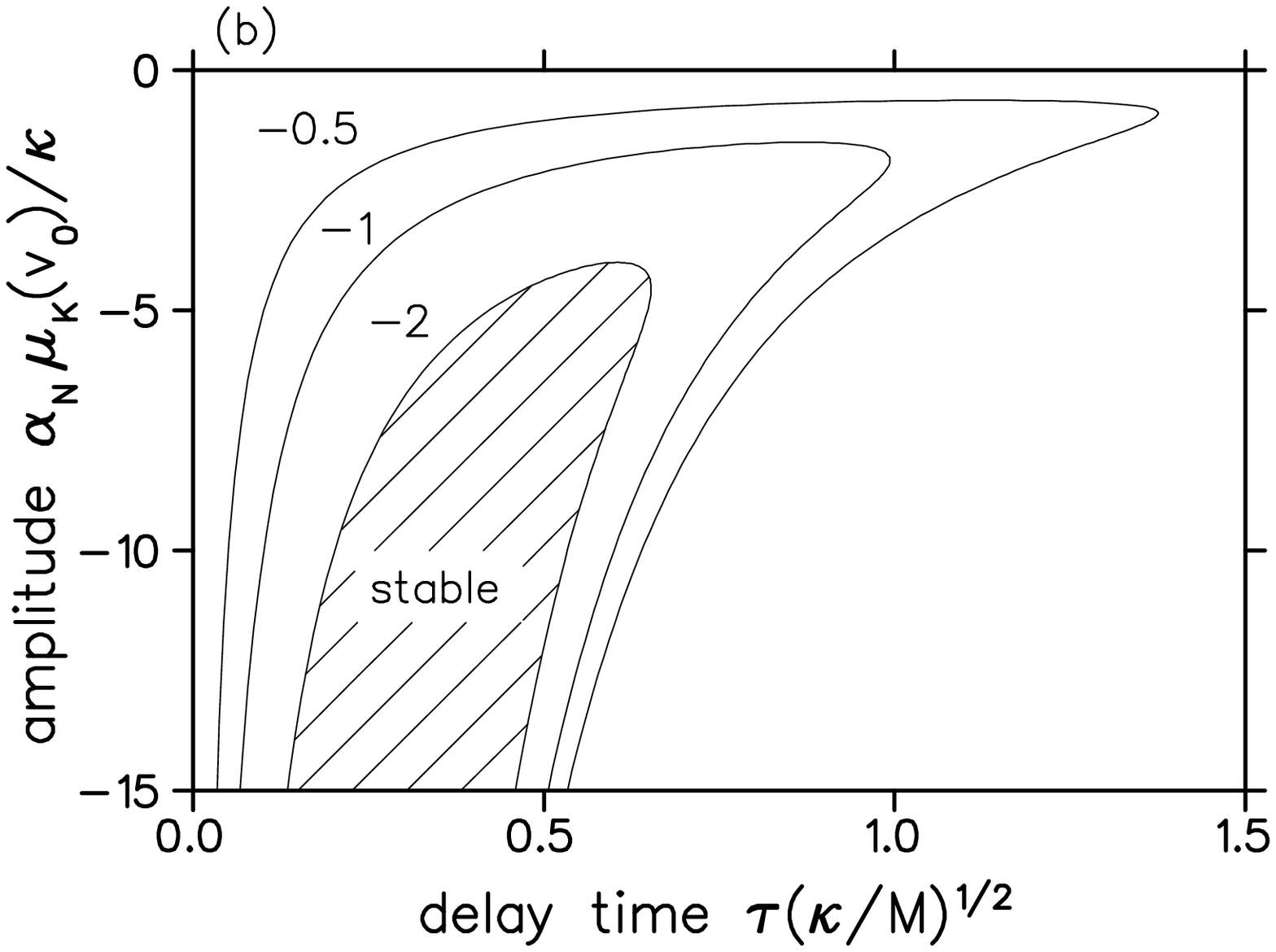}\vspace{3mm}
\caption[fstab]{\protect\label{f.stab}Boundaries of stable continuous
sliding for (a) pure velocity control (i.e., $\alpha_N=0$) and (b)
pure load control (i.e., $\alpha_v=0$). Different boundaries are
denoted by the values of $\mu_K'(v_0)N_0/\sqrt{\kappa M}$.
}
\end{figure}

Figure~\ref{f.stab} shows that unstable continuous sliding states can
be stabilized with both types of control.  Because of the
multiplicity of the solutions of (\ref{tc}) we also get stability
regions for larger values of $\tau$. But there are less interesting
because of larger time scales on which the system relaxes into the
continuous sliding state. Pure velocity control (i.e., $\alpha_N=0$)
is less efficient than load control because the stability area is
smaller and it disappears if $\mu_K'(v_0)N_0\lesssim
-0.68\sqrt{\kappa M}$. 

\begin{figure}
\epsfxsize=80mm\epsffile{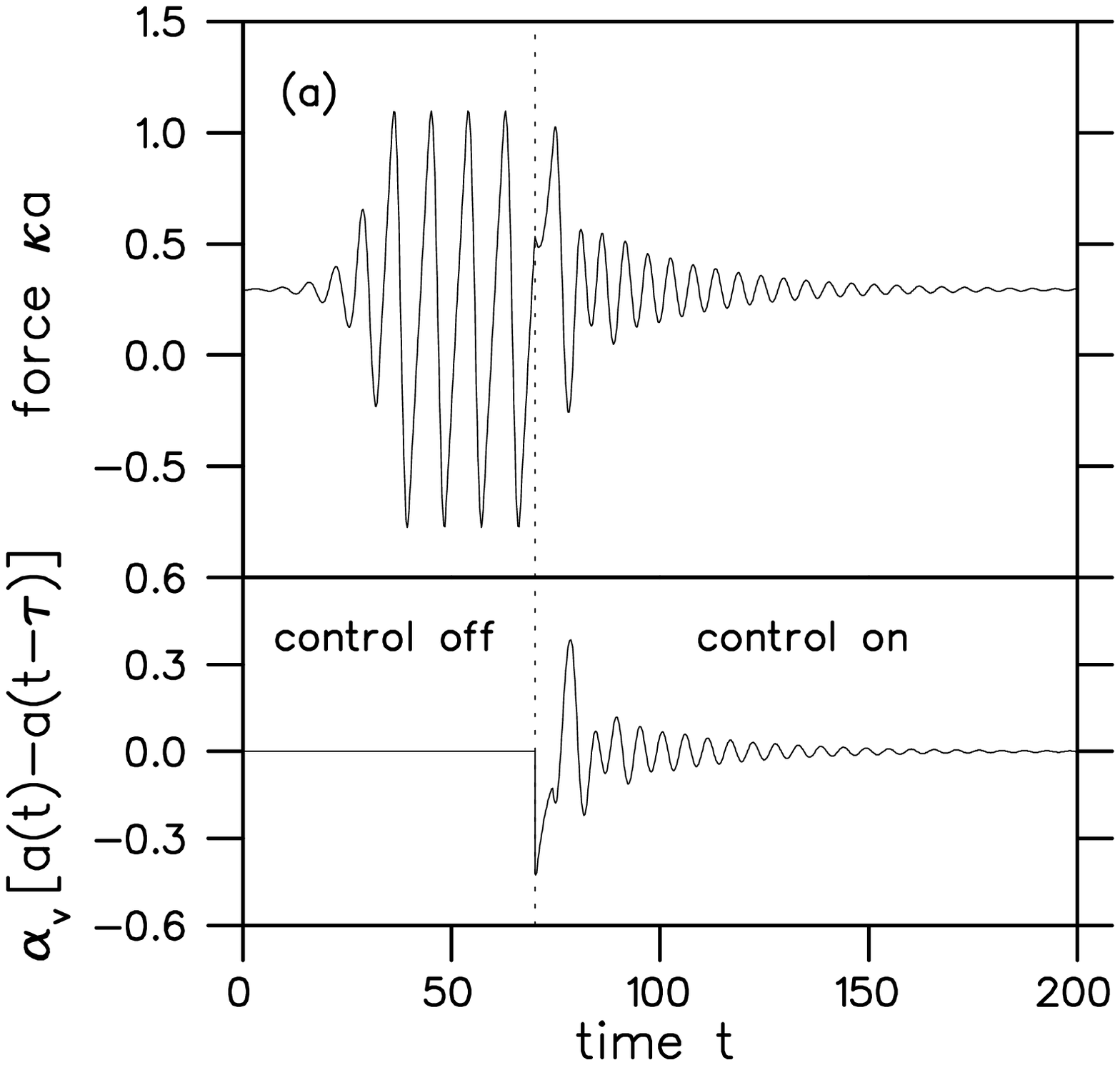}\vspace{3mm}
\epsfxsize=80mm\epsffile{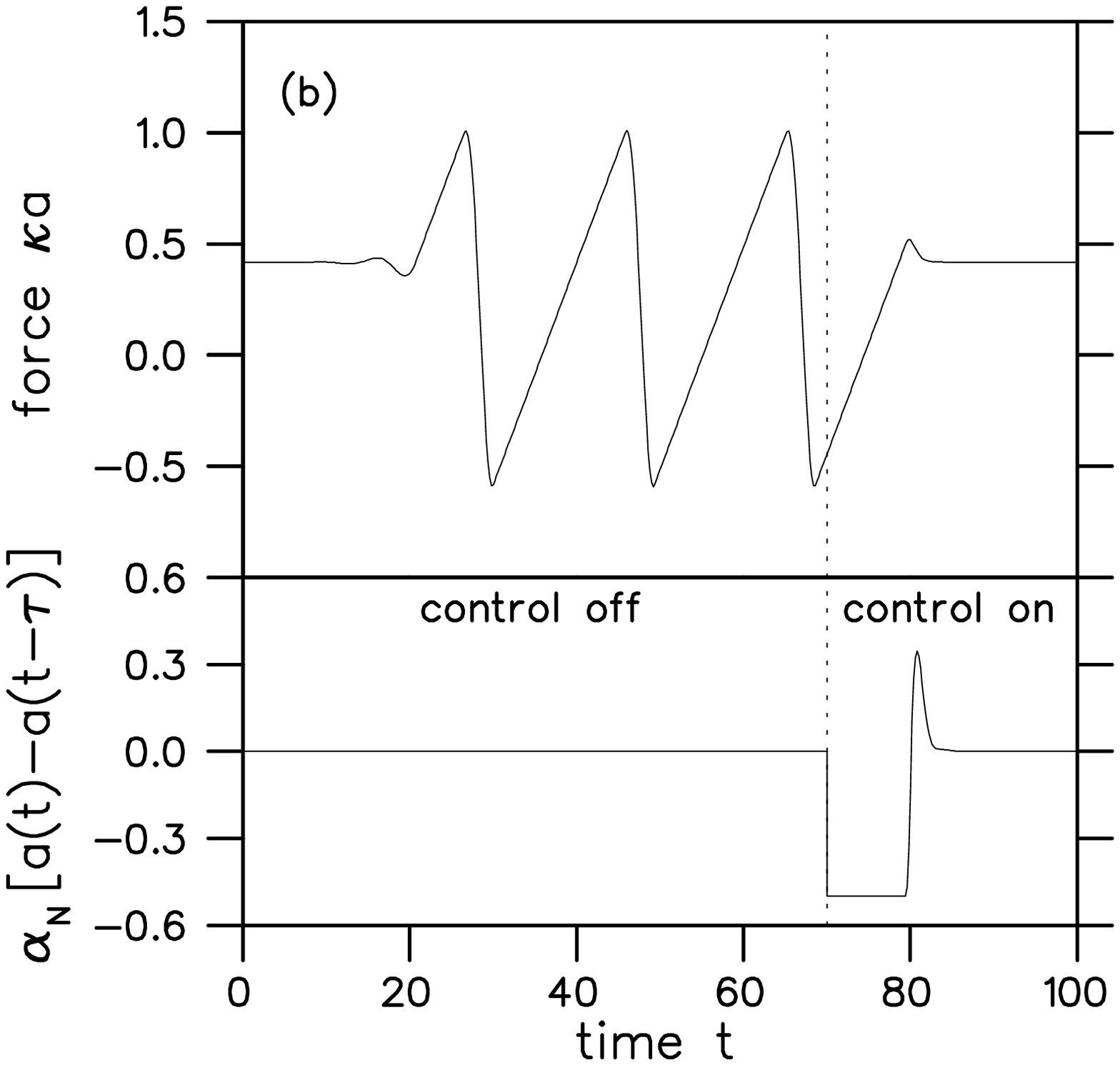}\vspace{3mm}
\caption[fcs]{\protect\label{f.cs}Turning stick-slip motion into
continuous sliding.  The simulations of Eq.~(\ref{eqm}) were done
with the purely velocity-weakening friction law $\mu_K(\dot
x)=\mu_0/(1+\dot x/v_s)$ with $\mu_0=0.5$ and $v_s=0.5$. The
simulation starts with an initial state near the unstable continuous
sliding state. The instability leads to stick-slip motion. At $t=70$
the control was switch on [velocity control in case (a) and load
control in case (b)]. The parameters are $M=\kappa=\mu_S=N_0=1$, (a)
$v_0=0.35$, $\alpha_v=-0.33$, $\alpha_N=0$, $\tau=4.15$, and (b)
$v_0=0.1$, $\alpha_v=0$, $\alpha_N=-10$, $\tau=0.5$.
}
\end{figure}

In pure load control there is always a stable region. Only the
minimal value of $|\alpha_N|$ increases with increasing $|\mu_K'|$.
For $\tau\to 0$ a more simple formula for the stability boundary can
be given. For small delay times (i.e., $\tau\ll \sqrt{M/\kappa}$) one
can approximate $a(t)-a(t-\tau)$ by the time derivative of $a$. Hence
the equation of motion becomes a differential equation and (\ref{cp})
becomes a second-order polynomial since
$(1-e^{-\lambda\tau})\to\lambda\tau$. It is easy to see that pure
velocity control (i.e., $\alpha_N=0$) is not able to stabilize
unstable continuous sliding states whereas for pure load control
(i.e., $\alpha_v=0$) the stabilization works if 
\begin{equation}
  \alpha_N<\frac{\mu_K'(v_0)N_0}{\mu_K(v_0)\tau}.
  \label{anc2}
\end{equation}

We have confirmed our analytical results by numerical simulations of
the equation of motions. Furthermore they show that the basin of
attraction of a formerly unstable continuous sliding state can be quite
large. The examples in figure~\ref{f.cs} show that stick-slip
oscillations disappear after the control is switched on. In our
simulations we found that stick-slip motions survives if the applied
velocity $v_0$ is below some critical value $v_c$.  Again pure
velocity control is less robust than load control. For example,
stick-slip motion can not be destroyed by pure velocity control for
that value of $v_0$ for which in Fig.~\ref{f.cs}(b) load control turns
easily stick-slip motion into continuous sliding.  As a rule of thumb
we found that in the case of pure velocity control the sticking time
has to be of the same order or less than the slipping time.  

\begin{figure}
\epsfxsize=80mm\epsffile{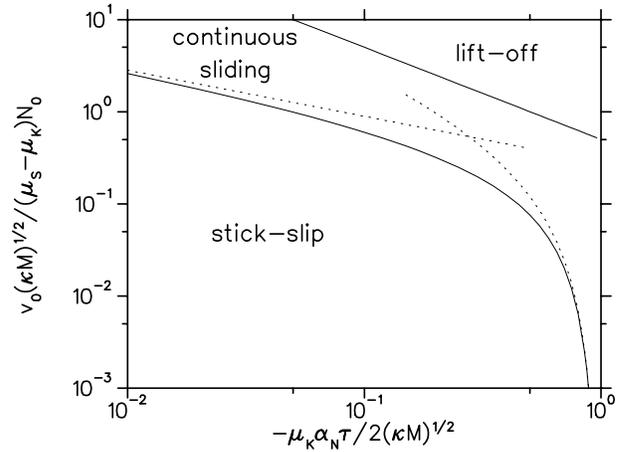}\vspace{3mm}
\caption[fsscs]{\protect\label{f.sscs}Boundary of stick-slip motion.
The lower solid curve is $v_c$ for $\tau\to 0$ and Coulomb's law
(i.e., $\mu_K=const<\mu_S$). The dotted curves are analytic
approximations (for more details, see the main text). 
The upper solid curve is the lift-off condition. 
The ratio of both friction coefficients is $\mu_K/\mu_S=0.5$.
}
\end{figure}

In the case of load control it is possible to turn stick-slip motion
into continuous sliding for arbitrary small values of $v_0$. To see
this we discuss an analytically treatable case. We assume that 
$\mu_K$ does not depend on the sliding velocity (Coulomb's law). 
Furthermore we restrict ourself to the limit $\tau\to 0$ (but
$\alpha_N\tau$ finite). Both assumptions turn the equation of motion
into a linear differential equation for the slip motion:
\begin{equation}
  M\frac{d^2\dot x}{dt^2}-\mu_K\alpha_N\tau\frac{d\dot x}{dt}
   +\kappa(\dot x-v_0)=0,
  \label{eqm.l}
\end{equation}
with the initial conditions
\begin{equation}
  \dot x(0)=0,\quad
  \frac{d\dot x}{dt}(0)=\frac{\mu_S-\mu_K}{M}(N_0+\alpha_N\tau v_0),
  \label{ic.l}
\end{equation}
assuming that the slip motion is just starting at $t=0$. The continuous
sliding state is stable for $\alpha_N<0$. It will be approached in the
long-time limit if $\dot x(t)>0$, for $t>0$ (i.e., no resticking).
For $\alpha_N\tau<-2\sqrt{\kappa M}/\mu_K$, Eq.~(\ref{eqm.l})
describes an overdamped harmonic oscillator. Thus the solution
$\dot x(t)$ increases monotonically and never resticks. Therefore
stick-slip motion disappears even for infinitesimal small $v_0$, i.e.,
$v_c=0$.  In the underdamped case one gets resticking if $v_0<v_c$.
To calculate $v_c$ one has to solve $\dot x(T)=\ddot x(T)=0$, with
$T>0$. This can be done numerically. The result for $\mu_K=0.5\mu_S$
is shown in figure~\ref{f.sscs}. In the limits $\alpha_N\to 0$ and
$\alpha_N\to -2\sqrt{\kappa M}/\mu_K\tau$ we get the approximations
$v_c=\tilde{v}/2\sqrt{\pi\gamma}$ and $v_c=\tilde{v}
\gamma^{-1}\exp\bigl(-1-\pi\gamma/\sqrt{1-\gamma^2}\bigr)$, resp.,
where $\tilde{v}\equiv(\mu_S-\mu_K)N_0/\sqrt{\kappa M}$ and
$\gamma\equiv -\mu_K\alpha_N\tau/2\sqrt{\kappa M}$.  Note that $v_0$
has to be less than $N_0/(-\alpha_N\tau)$ otherwise the control
mechanism would lead to a lift-off in the sticking phase. It is easy
to show that the lift-off curve is always above $v_c$ (see
Fig.~\ref{f.sscs}).

Recently Rozman {\em et al.\/} proposed a different method for
stabilizing the continuous sliding state \cite{roz.97}. Their method
is similar to the method of Ott {\em et al.\/} for stabilizing
periodic orbits in a chaotic attractor \cite{ott.90}. For this method
one has to reconstruct the Poincar\'e return map near the unstable
orbit.  This is done by observing the system dynamics without
control. Rozman {\em et al.\/} used the normal
force as the parameter of controlling.  
 
Compared with our method the advantage of the method of Rozman {\em
et al.\/} is that one has not to rely on macroscopic equations of motion
like (\ref{eqm}). Such equations of motion are reliable for large sliding
velocities but it is well-known that they may be not correct for
small velocities, especially in the case of transitions from sticking
to sliding and vice versa \cite{elm.97,rui.83,bau.95,car.96,per.97}.
In fact Rozman {\em et al.\/} have tested their method for a simple
model where in addition to the macroscopic degree of freedom (i.e.,
the position of the sliding block) an internal degree of freedom
appears which describes the state of a lubricant. 

There are two disadvantages of the method of Rozman {\em et al.\/}.
First, controlling methods \`a la Ott {\em et al.\/} work only in
the vicinity of periodic orbits.  If these orbits are embedded in a
chaotic attractor the system will eventually come close to them.
Therefore, turning stick-slip motion into continuous sliding is
possible only, if the stick-slip motion is erratic enough to be close
to the continuous sliding state.  Otherwise, the method works only if
one starts at a large stage velocity where the continuous sliding state
is already stable and then slowly decreases the velocity below the value
where the continuous sliding state becomes unstable \cite{roz.97}. A
delayed-feedback method is not restricted to the vicinity of the 
unstable orbit. Of course starting far away from the orbit may 
lead to strong controlling forces
at the beginning (see Fig.~\ref{f.cs}). But they decay exponentially by 
approaching the orbit. The second disadvantage of the method of
Rozman {\em et al.} is the
necessity of reconstructing the dynamics. This may be more or less
difficult depending on the details of the dynamics of the internal
degrees of freedom at the friction interface.  For technological
applications this might be important especially because the
reconstruction has to be recalibrated from time to time.

In this Letter we have introduced two robust methods of stabilizing
continuous sliding. They are also able to destroy regular stick-slip
motion. Both methods rely on a delayed feedback where the feedback
variable is the elastic deformation of the sliding bodies or the
machinery. The controlling parameter is either the applied velocity
or the normal load. The velocity control is less robust than the load
control. 

There are two fields of application of controlling friction.
Obviously there will be technological applications for reducing 
vibration and wear. But controlling friction experiments can also be
used to increase our understanding of the physics of dry friction.
For example, using these methods one can measure the effective
friction force as a function of the sliding velocity even in the
velocity-weakening regime.

\acknowledgments
I am gratefully acknowledge valuable discussions with Y. Klafter, M.
Rozman, and M. Urbakh.


\begin{references}
\bibitem{rem1}
Note that a
velocity-strengthening friction law does not mean the absence of
stick-slip motion. There is often a large regime of bistability
\cite{elm.97}.
\end{references}
\end{document}